%% file: document.tex
\newif\ifisdraft
\isdraftfalse  % draft = comments
\documentclass[runningheads]{llncs}

\usepackage{graphicx}

\usepackage{cite}
\usepackage{amsmath,amssymb,amsfonts}
\usepackage{algorithmic}
\usepackage{textcomp}
\usepackage{xcolor}
\usepackage[hidelinks]{hyperref}
\usepackage{mdframed}
\usepackage{mathtools}
\usepackage{array}
\usepackage{float}
\usepackage{mathpartir}
\usepackage{amsmath}
\usepackage{todo}
\usepackage{color}
\usepackage{listings}
\usepackage{subcaption}

\definecolor{comment}{rgb}{0.25,0.5,0.375}
\definecolor{keyword}{rgb}{0.5,0,0.33}
\definecolor{string}{rgb}{0.15,0.0,1.0}

\newcommand{\sem}[1]{[\![#1 ]\!]}

\lstdefinelanguage{csx}
{
	morekeywords={type, int, bool, enum, list, forall, and, action, parameter, param, device, location, component, value, scenario, test, succeeds, fails, minimize, maximize, if, else, config},
	literate=*{->}{$\rightarrow$~}{2}
            {>=}{\tiny$\geq$}2
            {<=}{\tiny$\leq$}{1}
            {/\\}{$\wedge$}{2}
            {\\/}{$\vee$}{2}
            {*}{$\ast$}{1},
	morecomment=[l]{//},
    basicstyle=\fontsize{8}{9.6}\ttfamily
}

\lstdefinelanguage{minizinc}
{
	morekeywords={enum, int, bool, array, var, of, constraint},
	literate=*{->}{$\rightarrow$~}{2}
            {>=}{$\geq$}{2}
            {<=}{$\leq$}{2}
            {/\\}{$\wedge$}{2}
            {\\/}{$\vee$}{2}
            {*}{$\ast$}{1},
	morecomment=[l]{\%},
	basicstyle=\fontsize{8}{9.6}\ttfamily
}

\mdfsetup{%
    innerleftmargin=5pt,
    innerrightmargin=5pt,
    innertopmargin=3pt,
    innerbottommargin=3pt
}

\begin{document}

\title{
    Configuration Space Exploration\\for Digital Printing Systems
    \thanks{
        \copyright~The Authors; licensed under Creative Commons License CC-BY.
        \href{https://doi.org/10.1007/978-3-030-92124-8_24}{https://doi.org/10.1007/978-3-030-92124-8\_24}.
        This version is an extended version with appendices on declarative semantics and inhabitance.
    }
}
%
%\titlerunning{Abbreviated paper title}
% If the paper title is too long for the running head, you can set
% an abbreviated paper title here
%
\author{Jasper Denkers\inst{1}\orcidID{0000-0003-3014-8324} \and
Marvin Brunner\inst{2} \and \\
Louis van Gool\inst{3} \and
Eelco Visser\inst{4}\orcidID{0000-0002-7384-3370}}
\authorrunning{J. Denkers et al.}
% First names are abbreviated in the running head.
% If there are more than two authors, 'et al.' is used.
%
\institute{
    Delft University of Technology,
    %Delft,
    the Netherlands,
    \email{j.denkers@tudelft.nl}
    \and
    Canon Production Printing B.V.,
    %Venlo,
    the Netherlands,
    \email{marvin.brunner@cpp.canon}
    \and
    Canon Production Printing B.V.,
    %Venlo,
    the Netherlands,
    \email{louis.vangool@cpp.canon}
    \and
    Delft University of Technology,
    %Delft,
    the Netherlands,
    \email{e.visser@tudelft.nl}
}
\maketitle              % typeset the header of the contribution

\begin{abstract}
\input{sections/00-abstract}
\end{abstract}

\input{body}

\section*{Acknowledgments}
We thank the reviewers for their feedback.
This research was partially supported by a grant from the Top Consortia for Knowledge and Innovation (TKIs) of the Dutch Ministry of Economic Affairs and by Canon Production Printing.
We thank Bas Hermus for providing a 3D drawing of perfect binding.
This work is related to the European patent application EP3855304 A1 which is published on 28 July 2021.

\bibliographystyle{splncs04}
\bibliography{manual,csx,dsls-in-practice}

\input{appendices}

\end{document}

%% file: sections/00-abstract.tex
Within the printing industry, much of the variety in printed applications comes from the variety in finishing.
Finishing comprises the processing of sheets of paper after being printed, e.g. to form books.
The configuration space of finishers, i.e. all possible configurations given the available features and hardware capabilities, are large.
Current control software minimally assists operators in finding useful configurations.
Using a classical modelling and integration approach to support a variety of configuration spaces is suboptimal with respect to operatability, development time, and maintenance burden.

In this paper, we explore the use of a modeling language for finishers to realize optimizing decision making over configuration parameters in a systematic way and to reduce development time by generating control software from models.

We present CSX, a domain-specific language for high-level declarative specification of finishers that supports specification of the configuration parameters and the automated exploration of the configuration space of finishers.
The language serves as an interface to constraint solving, i.e., we use low-level SMT constraint solving to find configurations for high-level specifications.
We present a denotational semantics that expresses a translation of CSX specifications to SMT constraints.
We describe the implementation of the CSX compiler and the CSX programming environment (IDE), which supports well-formedness checking, inhabitance checking, and interactive configuration space exploration.
We evaluate CSX by modelling two realistic finishers.
Benchmarks show that CSX has practical performance (\textless$1$s) for several scenarios of configuration space exploration.

\ifisdraft\pagebreak\fi

%% file: body.tex
\ifisdraft\clearpage\fi
\input{sections/01-introduction}
\ifisdraft\clearpage\fi

\input{sections/02-context}
\ifisdraft\clearpage\fi

\input{sections/03-csx}
\ifisdraft\clearpage\fi

\input{sections/04-semantics}
\ifisdraft\clearpage\fi

\input{sections/05-implementation}
\ifisdraft\clearpage\fi

\input{sections/06-evaluation}
\ifisdraft\clearpage\fi

\input{sections/07-related}
\ifisdraft\clearpage\fi

\input{sections/08-conclusions}
\ifisdraft\clearpage\fi

%% file: sections/01-introduction.tex
\section{Introduction}

Digital printing systems are flexible manufacturing systems, i.e. manufacturing systems that are capable of adjusting their abilities to manufacture different types and quantities of products, without expensive hardware changes.
The variety in printing applications stems from both printing (printing on sheets of paper) and finishing (processing collections of printed sheets, e.g. to form a book).
The \emph{configuration space} for a digital printing system consists of all possible configurations given the system's features and hardware constraints.
For producing a booklet of a particular size, a printed stack of sheets can be stitched, it can be folded, and it can be trimmed.
Optionally, the sheets can be rotated in an intermediate production step such that a single trimming component can be used for trimming in multiple dimensions.
The decisions made for these manufacturing parameters influence important factors such as productivity (production time increases when sheets are rotated) or efficiency (paper is wasted when input sheets are trimmed).

Ideally, control software assists operators in exploring the configuration space.
For example, given some available paper and the intent to produce a booklet, the software should automatically derive a viable manufacturing configuration.
Such a configuration e.g. comprises the orientation of the input sheets, the number of stitches, and the amount of side and face trimming needed to get the desired end result.
In addition, an optimization objective can be relevant while finding a configuration, e.g. minimizing paper waste.
The control software and user interfaces of state of the art digital printing systems do not support such automated configuration space exploration.
Instead, operators have to provide configurations for finishers manually.
A configuration can be simulated; by ``executing'' the finishing process in software, finishing viability can be checked without wasting resources.
Still, it remains a cognitively intensive task for operators to find a valid or optimal configuration.

Finishers are produced by many vendors and integrating them with printers is non-trivial.
Such integration involves connecting the control software of the printer and finishers and driving embedded software components.
Using a classical modeling and integration approach to support the variety of finishing is suboptimal with respect to development time and maintenance burden.
Issues with such a classical approach are the long code-build-test cycle and the large amount of finisher vendors and models that must be supported for many years.
The translation of the mechanical specifications into control software code gives rise to additional complexity.

Our objective is to obtain an effective, efficient, and scalable method for modeling finishers and obtaining control software for finishers that support automated configuration space exploration.
In this work, we investigate how linguistic abstraction can help to model the configuration space of digital printing systems, and how we can automatically derive environments for configuration space exploration from such specifications.

The global characteristics of finishers make the use of constraint (SMT) solving a natural fit for realizing environments for configuration space exploration.
For example, trimming the paper along a certain dimension might impose a specific orientation or transformation in an earlier production step.
A constraint-based approach considers its specifications as global and will take into account interdependent system-level constraints when finding solutions, i.e., configurations.
A constraint-based model of a finisher contains a representation of the input materials at intermediate locations in the system.
However, for modelling domain objects such as sheets and stacks, abstraction mechanisms such as classes are not naturally available in SMT modelling.
An SMT model of a finisher requires low-level encoding of the properties of the materials at all locations.
Therefore, expressing finishers in SMT by hand is tedious, error prone, and is not in terms of domain concepts.
Additionally, an SMT model of a finisher is complex to understand and difficult to maintain.

In this paper, we present CSX, a domain-specific language for the high-level declarative specification of finishers.
The language supports specification of input materials, configuration parameters, output products, and finishing constraints in terms of domain concepts.
The CSX IDE supports the development and checking of specifications and the automated derivation of an environment for configuration space exploration by operators of the finishers.

CSX provides a domain-specific interface to SMT solving by abstracting and structuring over low-level properties.
We translate specifications to the SMT domain and use existing solvers to find solutions at the level of properties and finishing parameters.
A solution in the SMT domain corresponds to a valid configuration.
Unsatisfiability at the SMT level indicates an empty configuration space, i.e., no finishing possibilities.
By mapping SMT solutions back to the specification level, we can interpret CSX specifications in multiple modes: checking whether a configuration is valid, finding an (optimal) configuration, and validating specifications.
By caching invocations of the solver in the IDE, response times are improved which leads to an interactive editing experience.

The approach of specifying a finisher with CSX and deriving control software has similarities with the approach of simulation in control software.
Both approaches take representations of the products being produced at intermediate locations in the devices.
However, while simulation involves an operational and sequential application of transformations on objects, a constraint-based approach considers the devices globally.
CSX improves over simulation in the sense that it derives environments that can search for (optimal) configurations in an automated way, taking system-global interdependencies into account.

We evaluated the design and implementation of CSX by modelling two finishers: a perfect binder and a booklet maker.
In the process of modelling these devices, we have experimented with various encodings.
For both cases, we benchmark the configuration space exploration performance for several scenarios.

\paragraph*{Contributions}

To summarize, the contributions of this paper are the following:

\begin{itemize}

  \item We have developed CSX, a declarative language for the specification of finishers at the conceptual level of the domain. We interpret CSX specifications for several modes of configuration space exploraton: checking whether configurations are valid, finding optimal configurations under objectives, and interactively validating specifications.
  
  \item We define a denotational semantics of CSX in terms of SMT constraints that serves as an interface to solvers that can be used to find models in order to check inhabitance of a specification and to explore the configuration space of the specified finisher.

  \item We realize a programming environment for CSX that integrates an SMT solver as back-end and that presents solutions in terms of the specification.
  
  \item We evaluate CSX by specifying two types of finishers: a perfect binder and a booklet maker. For these cases, we benchmark the performance for a configuration space exploration scenario with and without optimization.

\end{itemize}

%% file: sections/02-context.tex
\section{Finishers in the Digital Printing Domain}
\label{sec:context}

In this section, we discuss the domain of digital printing systems with finishers.
Complete printing systems for e.g. producing books include, in addition to printing itself, finishing capabilities.
Finishing comprises the processing of printed sheets of paper into end products.
For example, a stack of printed sheets could be stapled, folded, and trimmed to result into a booklet; stapling, folding, and trimming are finishing operations.
Finishing devices need to be integrated with the printing system for realizing an integrated end-to-end experience for the print system end-users (i.e. operators in print shops).

The turnaround time of integrating finishers with printers is high because of multiple challenging aspects.
First, finishers are often produced by external vendors and communication is mostly documentation based and thus requires interpretation, reviews, implementation, and testing.
Second, obtaining good system behavior requires mechanical, electrical and software interfaces to be matched well between the printer and finisher.
Third, total aspects such as reliability are the result of all the mentioned interfaces to be well designed.
Considerable testing time is needed to confirm reliability.

Creating control software that is user-friendly for operators is difficult and requires a lot of manual programming.
This is because of the high variability and many configuration parameters in print and finishing systems.
A typical print and finishing system has more than 200 accessible parameters for the operator, that are also interdependent.
Because the whole production process is a sequence of production steps, choices that you have to make in the beginning influence the steps later on.
From the product line perspective, the control software supports tens of different finisher types, that each of them can have more than 100 commercial variations.
For all variations, the parameters that are accessible for operators can vary.

Ideally, operators can use the combination of a printer with finishers as an end-to-end solution instead of having to configure each device separately.
Additionally, optimization capabilities are also useful when considering the system as a whole.
For example, an operator would like to produce booklets with the available resources and while minimizing paper waste or while optimizing productivity.
If the different configuration possibilities impose a tradeoff between e.g. resource consumption and productivity, an operator should be able to make a motivated choice with ease, i.e., without thinking about and manually trying out many combinations of configuration parameters.

\subsection{Perfect Binding}

\input{figures/perfect-binding}

As an example, we discuss a \emph{perfect binder}: a finisher that produces books by binding a stack of sheets with glue and by covering the bookblock in a cover sheet.
A perfect binder typically has two inputs: one for the stack of sheets that form the book block and one for the cover sheets.
Figure~\ref{fig:perfect-binding} shows the perfect binding process.
Figure~\ref{fig:perfect-binding-topview} depicts the components of a perfectly bound book, viewed from above.

After collecting a stack of sheets, jogging makes sure the stack of sheets becomes aligned in a corner of the spine.
Then, a clamp grasps the bookblock under pressure.
Next, a few millimeters of paper are milled along the spine edge to prepare the spine for application of glue.
Milling makes the paper along the spine rough, improving adherence of the glue.
Then, the spine travels through a bath of heated glue.

Separately, cover sheets are prepared before being bound around the bookblock.
The preparation consists of creasing, i.e., applying pressure on the paper to ease folding of the paper later.
Two creases are applied at the location of the cover that end up along the edges of the spine of the book.
These creases improve the fit of the cover along the spine of the book block, supporting a tight fold around the spine.
Additionally, two courtesy creases are applied on the cover.
Courtesy creases are applied on the front and back of the resulting book to support the folding of the cover sheet.
Note that courtesy creases are applied at the opposite side as the spine creases, as they are used for folds in opposite directions.

After preparing the bookblock and cover, the covering occurs.
The bookblock with glue is positioned in the center of the cover sheet.
The cover sheet is folded around the bookblock and fixed with a clamp.
After a delay for the glue to solidify, the book is released.
In practice, the resulting book could be processed further in a cutting machine to trim along the edges of the book and cover to result into a nice book.

Perfect binders are flexible in the books they can produce, e.g. in terms of sheet size or book thickness.
Not all flexible manufacturing steps have impact on the configuration space.
For example, jogging and glueing occur automatically and are configured by the device itself based on measurements.
Other settings such as the milling depth and positioning of the bookblock on the cover are of interest to the operator and therefore do impact the configuration space; e.g. more milling might increase the overall production time.

%% file: figures/perfect-binding.tex
\begin{figure}[t]
	\centering

	\minipage{0.4\columnwidth}
		\includegraphics[width=\columnwidth]{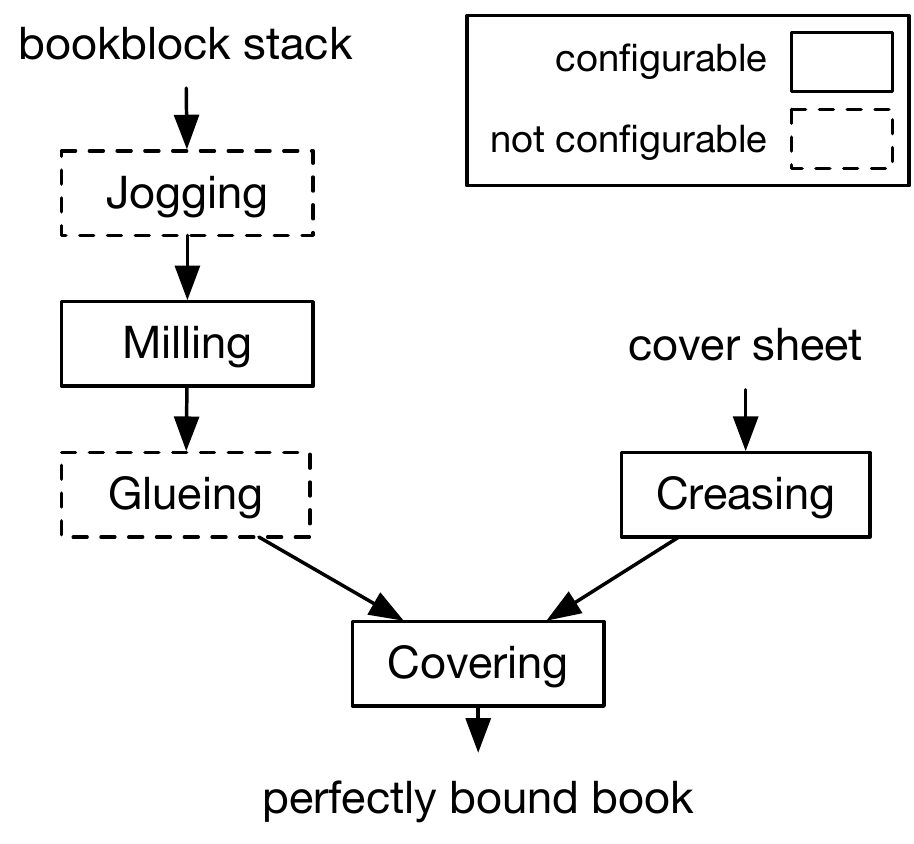}
		\caption{
			Schematic view of the perfect binding book producing process.
			Only milling, creasing, and covering are configurable and therefore impact the configuration space.
			Jogging and glueing are automatically configured by the device itself.
		}
		\label{fig:perfect-binding}
	\endminipage\hfill
	\minipage{0.4\columnwidth}
		\includegraphics[width=\columnwidth]{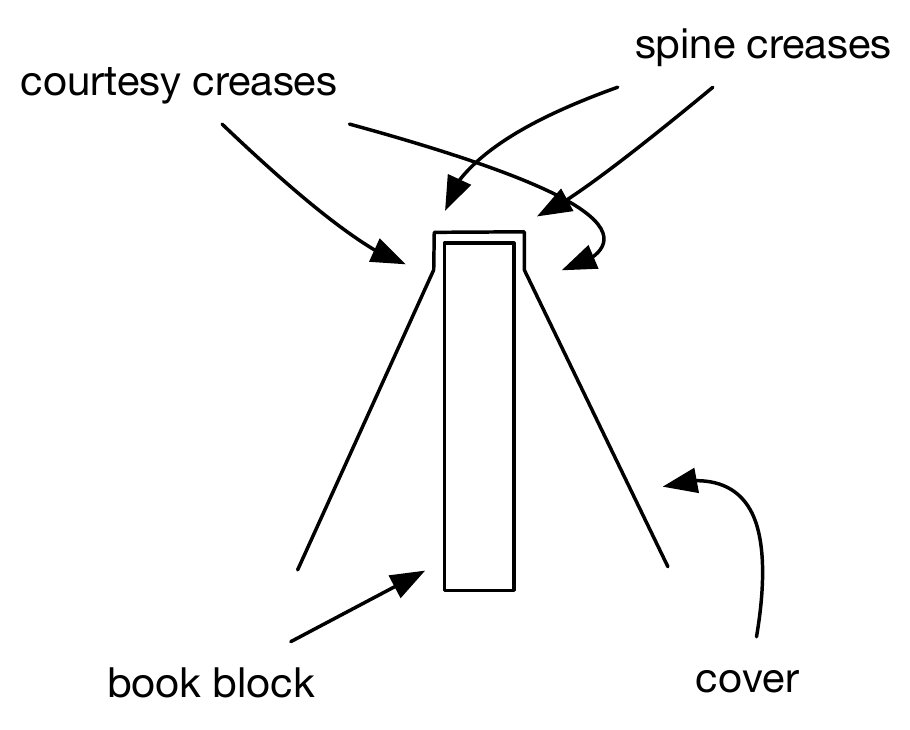}
		\caption{
			A perfectly bound book viewed from the top.
			Spine creases result into a sharper fold, reduce wrinkles, and improve the fit of the cover around the bookblock.
			Courtesy creases ease opening the front and back part of the cover.
			Glue in the spine holds the bookblock sheets and cover together.
		}
		\label{fig:perfect-binding-topview}
	\endminipage
\end{figure}

%% file: sections/03-csx.tex
\section{CSX}
\label{sec:csx}

\input{figures/perfect-binding-layout}
\input{figures/perfect-binding-spec-types}
\input{figures/perfect-binding-spec-actions}
\input{figures/perfect-binding-spec-device}

The key idea of CSX is that we model objects such as sheets and stacks and that we specify symbolic values, i.e. instances, for these objects at several intermediate steps in the finishing process.
By adding constraints and indicating configuration parameters, a specification defines the configuration space of a device.
In CSX we also describe \emph{jobs}, i.e., (partial) descriptions of the production process in terms of the production objects and parameters.
We achieve configuration space exploration by synthesizing configurations from a configuration space for a given job.

CSX is declarative: a specification in the language describes behavior and configuration spaces of finishers.
A CSX specification does not describe algorithms to compute configurations.
Specifications include relations between objects at locations in the systems.
We use the language to model devices as sequences of components that perform actions.
Components instantiate generic, reusable actions.
Actions establish a relationship between snapshots of objects in the finishers and thus, transitively, devices define a relation between all snapshots of the products being produced.
Parameters in actions represent a dimension of configuration that is of interest to operators of the devices.
Constraints restrict instances of types and restrict the behavior of actions and devices, reducing the configuration space.
We will now introduce the language concepts in more detail based on a specification for an example perfect binder such as described in Section~\ref{sec:context}.

Defined \emph{types} are records of properties that model objects at locations in a device.
In Figure~\ref{fig:perfect-binding-spec-types}, we define several types for the example perfect binder.
Dimensions (widths, heights, lengths, distances) are modelled with integers with a precision of 0.1mm, such that an integer value of $10$ stands for a length of $1$mm.
Types contain \emph{defining properties} that are of a primitive type (boolean or integer) or of a defined type such that types can be nested.
The nesting of types may not contain a cycle.
Types optionally contain \emph{constraints} and \emph{derived properties}.
Constraints restrict the inhabitants of a type.
In Figure~\ref{fig:perfect-binding-spec-types}, the constraints (between square brackets) e.g. restrict sheets to have positive non-zero width and height.
Derived properties are shorthands for expressions over other properties.
Defining properties are required to instantiate a type.
Derived properties are not required to instantiate a type and their values can be derived from other properties.
A derived property expression may refer to the type's properties and to other derived properties, but derived properties may not contain cyclic references.
In Figure~\ref{fig:perfect-binding-spec-types}, \texttt{Stack} has a derived property \texttt{volume} which is defined in terms of defining properties.

\emph{Actions} define a relation between locations.
In Figure~\ref{fig:perfect-binding-spec-actions}, we define several actions for the example perfect binder.
The body of an action definition contains parameters and constraints that indicate the relations between its parameters.

\emph{Devices} are sequences of \emph{components} connected through \emph{locations}.
Components instantiate actions and can restrict or specify behavior further by adding constraints.
Thus, action behavior is defined seperately from specific instantiations in components.
Therefore, actions are generic and potentially reusable between different device specifications.
Limitations of a particular instance of an action in a device can be specified by adding constraints to the component.
In Figure~\ref{fig:perfect-binding-spec-device} we define a perfect binder device by instantiating several actions in components and by connecting them through the locations.

\subsection{Configurations and Jobs}

A configuration for a device is a value assignment to all locations and parameters.
A valid configuration is a configuration that conforms to the constraints of the types of the locations, the actions, the components, and the device itself.
In practice, an operator is only interested in the values for the input and output locations, and not in the intermediate locations.

A job is an expression of intent for which a configuration needs to be found.
Whereas configurations are a complete specification of locations and parameters, we could see jobs as a partial configuration.
For example, a job could define the input and the output of the finisher.
The remaining parts of the configuration, i.e. the finishing parameters, need to be derived in order to instruct the finisher to realize the intent of the job.
Different usage scenarios of a device lead to different jobs and approaches to configuration.

\subsection{Exploration and Validation}
\label{sec:exploration-validation}

The CSX language supports configuration space exploration, which includes leveraging exploration at the specification level for validation.
Given the specification of a device, the language supports describing scenarios for testing devices by asserting expectations on configuration spaces.

The following test scenario validates that the correct cover dimensions are chosen for a particular input bookblock and desired output perfectly bound book:

\begin{lstlisting}[
    language=csx,
    basicstyle=\fontsize{7}{8.4}\ttfamily
]
scenario device ExamplePerfectBinder
 config bookIn = Stack(2125,2970,50)
 config out = PerfectBoundBook(Stack(2100,2970,50), Sheet(2100,2970), Sheet(2100,2970)) {
  [coverIn.width == 2100 + 2100 + 50]
  [coverIn.height == 2970]
  [toMill.millingDepth == 25]
}
\end{lstlisting}

The body of the scenario contains expectations (between square brackets) on its configuration space.
In particular, it validates the cover dimensions that must be chosen.
Since the configuration space could contain multiple configurations, expectations should only validate common properties of the configuration space and not on individual configurations.

Scenarios can optionally specify an objective.
\emph{Objectives} indicate a dimension for optimization of a property of the system, typically expressed using derived properties.
Potentially relevant objectives are e.g. maximizing throughput, minimizing energy consumption, or minimizing resource waste.
Alternatively, scenarios with optimization can characterize the device.
For example, based on the following scenario a scenario can be found for the largest book that the perfect binder can produce:

\begin{lstlisting}[
    language=csx,
    basicstyle=\fontsize{8}{9.6}\ttfamily
]
scenario device ExamplePerfectBinder
         maximize out.book.volume
\end{lstlisting}

%% file: figures/perfect-binding-layout.tex
\begin{figure}[t]
	\centering
	\includegraphics[width=0.6\columnwidth]{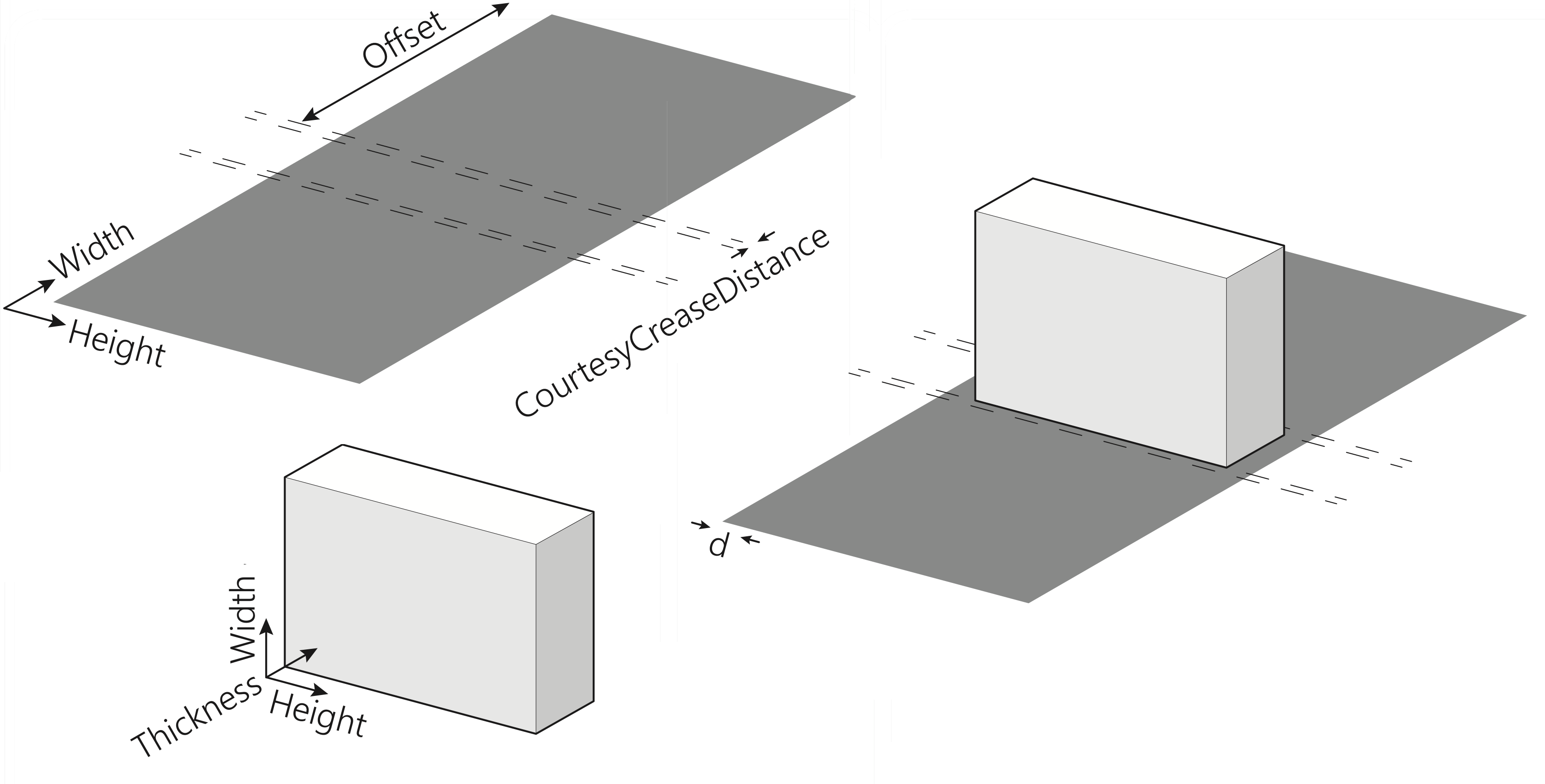}
	\caption{
		Components of a perfectly bound book (cover and bookblock) and the dimensions as how we use them in the CSX specification.
	}
	\label{fig:perfect-binding-layout}
\end{figure}

%% file: figures/perfect-binding-spec-types.tex
\begin{figure}
    \begin{minipage}{.475\columnwidth}
\begin{lstlisting}[
            language=csx
        ]
type Sheet {
 width: int, [width > 0],
 height: int, [height > 0]
}
type Stack {
 width: int, [width > 0],
 height: int, [height > 0],
 thickness: int, [thickness > 0],
 volume = width * height * thickness
}
type PerfectBoundBook {
 book: Stack,
 frontCover: Sheet, backCover: Sheet
}
\end{lstlisting}
    \end{minipage}\hfill
    \begin{minipage}{.475\columnwidth}
\begin{lstlisting}[
            language=csx
        ]
type CreasedSheet {
 sheet: Sheet,
 spineFront: Crease,
 spineBack: Crease,
 courtesyFront: Crease,
 courtesyBack: Crease
}

type Crease {
 // Offset:
 off: int, [off >= 0],
 // Direction: 0 = down, 1 = up
 dir: int, [dir == 0 or dir == 1]
}
\end{lstlisting}
    \end{minipage}

    \caption{
        The specification of types for the example perfect binder in CSX.
        Dimensions are in 0.1mm.
    }
    \label{fig:perfect-binding-spec-types}
\end{figure}

%% file: figures/perfect-binding-spec-actions.tex
\begin{figure}
\begin{lstlisting}[
        language=csx
    ]
action ToMill(in: Stack, out: Stack) {
 parameter millingDepth: int
 [millingDepth >= 0] [out.width == in.width - millingDepth]
 [out.height == in.height] [out.thickness == in.thickness]
}
action ToCrease(in: Sheet, out: CreasedSheet) {
 [out.sheet.width == in.width] [out.sheet.height == in.height]
 [out.spineFront.dir == 0] [out.spineBack.dir == 0]
 [out.courtesyFront.dir == 1] [out.courtesyBack.dir == 1]
 // Ensure a minimum distance between creases
 parameter minDist: int [minDist > 0] [out.courtesyFront.off >= minDist]
 [out.spineFront.off >= out.courtesyFront.off + minDist]
 [out.spineBack.off >= out.spineFront.off + minDist]
 [out.courtesyBack.off >= out.spineBack.off + minDist]
 [out.courtesyBack.off <= in.width - minDist]
}
action ToCover(cover: CreasedSheet, book: Stack, out: PerfectBoundBook) {
 parameter d: int [0 <= d and d <= cover.sheet.height - book.height]
 [out.book.width == book.width] [out.book.height == book.height]
 [out.book.thickness == book.thickness]
 [out.frontCover.width * 2 == cover.sheet.width - book.thickness]
 [out.frontCover.height == cover.sheet.height]
 [out.backCover.width * 2 == cover.sheet.width - book.thickness]
 [out.backCover.height == cover.sheet.height]
 [cover.spineFront.off * 2 == cover.sheet.width - book.thickness]
 [cover.spineBack.off * 2 == cover.sheet.width + book.thickness]
 parameter courtesyCreaseDist: int
 [cover.courtesyFront.off * 2 ==
   cover.sheet.width - book.thickness - courtesyCreaseDist * 2]
 [cover.courtesyBack.off  * 2 ==
   cover.sheet.width + book.thickness + courtesyCreaseDist * 2]
}
\end{lstlisting}
    \caption{
        The specification of actions for the example perfect binder in CSX.
        See Figure~\ref{fig:perfect-binding-layout} for the dimensions used in this specification.
    }
    \label{fig:perfect-binding-spec-actions}
\end{figure}

%% file: figures/perfect-binding-spec-device.tex
\begin{figure}
\begin{lstlisting}[
        language=csx
    ]
device ExamplePerfectBinder {
 location bookIn : Stack location coverIn : Sheet
 [1000 <= bookIn.height and bookIn.height <= 3000]
 [2000 <= bookIn.width and bookIn.width <= 5000]
 component toMill = ToMill(bookIn, milledBook) {
  [millingDepth <= 30] // Max 3mm of milling
  [bookIn.thickness < 170] // Max 17mm book thickness
 }
 location milledBook : Stack
 component toCrease = ToCrease(coverIn, creasedCover) {
  [minDist >= 50] // At least 5 mm between creases
 }
 location creasedCover : CreasedSheet
 component toCover = ToCover(creasedCover, milledBook, out) {}
 location out : PerfectBoundBook
}
\end{lstlisting}
    \caption{
        The specification of the example perfect binder device in CSX.
    }
    \label{fig:perfect-binding-spec-device}
\end{figure}

%% file: sections/04-semantics.tex
\input{figures/denotational-semantics}

\section{Denotational Semantics}
\label{sec:denotational-semantics}

Because of the declarative characteristic of CSX, a translation to SMT constraints is natural.
In this section, we define the denotational semantics of CSX that expresses a translation of CSX specifications to SMT constraints.
Figure~\ref{fig:denotational-semantics} contains the denotational semantics of CSX with the denotation expressed in MiniZinc~\cite{NethercoteSBBDT07, StuckeyFSTF14} definitions.
Because we use MiniZinc in the implementation of CSX (Section~\ref{sec:implementation}), we also use it as syntax for the denotation.
The MiniZinc grammar can be found online\footnote{\url{https://www.minizinc.org/doc-2.5.5/en/spec.html?highlight=grammar\#spec-grammar}}.

The intuition behind the translation is that the properties of locations and the parameters of components are mapped to constraint variables.
Additionally, all CSX-defined constraints translate to corresponding constraints in MiniZinc.
The translation is from the perspective of a device, making use of type and actions definitions of the CSX specification of which the device is part.

The translation starts with the \textsc{Device} rule, generating MiniZinc definitions for members of the device: locations, components, and device-level constraints.
The translation is defined under the context of a namespace $N$, starting with the empty namespace.
The naming scheme for constraint variables follow their corresponding hierarchical position in the CSX specification.
Since the translation is for a single device, we do not have to prefix the namespace with the device name.

A location translates into variables for its properties and into constraints to restrict its inhabitants (\textsc{Location}).
Locations are always of a user-defined type.
Each property of the type translates to variables.
If the property is of primitive type, the translation is a variable of this primitive type (\textsc{DefProp-PrimType)}).
If the property is of a user-defined type, the translation is the translation of its nested properties in the namespace of the property (\textsc{DefProp-DefType}).

The \textsc{Comp} rule defines the translation for a component, i.e. an action instantiation.
The action's parameters translate into variables in the namespace of the component (\textsc{Param}).
Both the action and the component can define constraints ($E_i^A$ and $E_i^C$, respectively).
These constraints are mapped to corresponding MiniZinc constraints.
Since the action's constraints are defined on the action's location parameters, and the action gets instantiated with specific location arguments, renaming is required.
The translation defines $R$: a mapping from the location's parameter names to the component's location argument names.
We only use the renaming for translating references to locations from constraints defined in the action definition.

The expressions that are used to define constraints, except references and projection, map mostly one-to-one to their MiniZinc counterparts.
For references and projection, we consider several cases.
A reference to property or parameter (\textsc{DefProp-Ref/Param-Ref}) translates to a name for $x$ in the context.
For example, a reference of $x$ in namespace $[a, b]$ will result in the denotation into a reference to name \texttt{a\_b\_x}.
For projection (\textsc{Proj}), we recursively translate the base expressions into a name and concetenate the projected name.

For a location reference, we consider two cases.
Location references from outside actions translate similarly as regular references (\textsc{Location-Ref}).
Location references within actions refer to location parameters, while the actions are instantiated with location arguments from a device.
Therefore, for such location references, we replace the location parameter name by the argument name for which it is instantiated (\textsc{ActionLocation-Ref}).

Types, actions, and devices can have derived properties.
These only translate into constraints if they are referenced, i.e. by replacing the reference with the body of the derived property and by propagating the namespace and location renaming (\textsc{DerProp-Ref}).
For the definition of derived properties, no translation takes place.
The definition of derived properties are ignored by $\dots$ in the specification.

Solutions found for the MiniZinc denotations are related to valid configurations for CSX specifications, and we can translate such solutions back to CSX specifations.
The correspondence between location properties and component parameters in CSX and MiniZinc is defined by the naming scheme used in the denotation, and mapping them back is thus straightforward.

%% file: figures/denotational-semantics.tex
\begin{figure*}
    
    \begin{mdframed}
        $\sem{ S' }_{S,N,R}$ = $M$ \hfill Specification part $S'$ of $S$ translates to $M$ in namespace $N$ with \\
        \hphantom{x} \hfill location renaming $R$ \\
        $N = [x_1, x_2, \dots, x_n]$ \hfill Namespace $N$ consisting of parts $x_1$ to $x_n$ \\
        $R = \{ \dots, L_i \rightarrow L'_i, \dots \}$ \hfill Renaming of location names $L_i$ to $L'_i$ \\
        $name([x_1, x_2, \dots, x_n]) = \texttt{$x_1$\_$x_2$\_$\dots$\_$x_n$}$ \hfill Identifier for namespace $[x_1, x_2, \dots, x_n]$ \\
        Locations $L$, components $C$, constraints $E$, defining properties $P$, types $T$, action parameters $PM$.
    \end{mdframed}

    \vspace*{-1.5\baselineskip}

    \begin{mdframed}[frametitle={Devices}]
        \vspace*{-2.5\baselineskip}
        \begin{mathpar}
            
            \sem {
                \texttt{device } d \texttt{ \{ }
                    L_1 \dots L_n,
                    C_1 \dots C_m,
                    E_1 \dots E_q,
                    \dots %DP_1 \dots DP_r
                \texttt{ \}}
            }_{S, [], \emptyset} =

        \end{mathpar}

        \vspace*{-2\baselineskip}
        
        \begin{mathpar}

            \bigcup_{i=1}^{n}
            \sem { L_i }_{S, [], \emptyset}
            \cup
            \bigcup_{i=1}^{m}
            \sem { C_i }_{S, [], \emptyset}
            \cup
            \bigcup_{i=1}^{q}
            \sem { E_i }_{S, [], \emptyset}
            \quad(\textsc{Device})

        \end{mathpar}
    \end{mdframed}

    \vspace*{-1.5\baselineskip}

    \begin{mdframed}[frametitle={Locations}]
        \vspace*{-2.25\baselineskip}
        \begin{mathpar}
        
            \inferrule{
                \texttt{type } T \texttt{ \{ }
                    P_1 \texttt{:} T_1 \dots P_n \texttt{:} T_n,
                    E_1 \dots E_m,
                    \dots
                \texttt{ \}} \in S
            }{
                \sem {
                    \texttt{location } L : T
                }_{S, [], \emptyset} =
                \bigcup_{i=1}^{n}
                \sem { P_i \texttt{:} T_i }_{S, [L], \emptyset}
                \cup
                \bigcup_{i=1}^{m}
                \sem { E_m  }_{S, [L], \emptyset}
            }
            \quad(\textsc{Location})
        \end{mathpar}

        \vspace*{-1.5\baselineskip}
        
        \begin{mathpar}
            \inferrule{
                T \in \{ \texttt{int}, \texttt{bool} \}
            }{
                \sem { P \texttt{:} T }_{S, N, \emptyset} =
                \texttt{var } T \texttt{ : } name(N +\!\!+ [P]) \texttt{ ; }
            }
            \quad(\textsc{DefProp-PrimType})
        \end{mathpar}

        \vspace*{-1.25\baselineskip}
        
        \begin{mathpar}
            \inferrule{
                \texttt{type } T \texttt{ \{ }
                    P_1 \texttt{:} T_1 \dots P_n \texttt{:} T_n,
                    E_1 \dots E_m,
                    \dots
                \texttt{ \}} \in S
            }{
                \sem {
                    P \texttt{:} T
                }_{S, N, \emptyset} =
                \bigcup_{i=1}^{n}
                \sem { P_n \texttt{:} T_n }_{S, N +\!\!+ [P], \emptyset}
                \cup
                \bigcup_{i=1}^{m}
                \sem { E_m }_{S, N +\!\!+ [P], \emptyset}
            }
            \quad(\textsc{DefProp-DefType})

        \end{mathpar}
    \end{mdframed}

    \vspace*{-1.5\baselineskip}

    \begin{mdframed}[frametitle={Components}]
        \vspace*{-3.25\baselineskip}
        \begin{mathpar}
        
            \inferrule{
                \texttt{action } A \texttt{(}
                    L_1 \texttt{:} T_1^L \dots L_n \texttt{:} T_n^L
                \texttt{)} \\ \texttt{\{}
                    \texttt{parameter } PM_1 : T_1^P \dots \texttt{parameter } PM_m : T_m^P,
                    E_1^A \dots E_q^A,
                    \dots
                \texttt{\}} \in S \\\\
                R = \{ L_1 \rightarrow L'_1, \dots, L_n \rightarrow L'_r \}
            }{
                \sem{ 
                    \texttt{component } C \texttt{ = } A \texttt{ ( }
                        L'_1 \dots L'_r
                    \texttt{ ) \{ }
                        E_1^C \dots E_s^C
                    \texttt{ \}}
                }_{S, [], \emptyset} = \\\\
                \bigcup_{i=1}^{m}
                \sem { \texttt{parameter } PM_m : T_m^P }_{S, [C], \emptyset}
                \cup
                \bigcup_{i=1}^{q}
                \sem {
                    E_i^A
                }_{S, [C], R}
                \cup
                \bigcup_{i=1}^{s}
                \sem {
                    E_s^C
                }_{S, [C], \emptyset}
            }
            \quad(\textsc{Comp})
        \end{mathpar}

        \vspace*{-1.5\baselineskip}
        
        \begin{mathpar}
            \inferrule{
                T \in \{ \texttt{int}, \texttt{bool} \}
            }{
                \sem { \texttt{parameter } PM \texttt{:} T }_{S, N, \emptyset} =
                \texttt{var } T \texttt{:} name(N $+\!\!+$ [PM]) \texttt{ ; }
            }
            \quad(\textsc{Param})

        \end{mathpar}
    \end{mdframed}

    \vspace*{-1.5\baselineskip}

    \begin{mdframed}[frametitle={Constraints \& References}]
        \vspace*{-1.5\baselineskip}
        
        \begin{mathpar}
            
            \sem{ \texttt{ [ } e \texttt{ ] } }_{S, N, R} = \texttt{constraint } \sem { e }_{S, N, R} \texttt{;}
            \quad(\textsc{Constraint})

        \end{mathpar}

        \vspace*{-2\baselineskip}

        \begin{mathpar}

            \inferrule{
                \text{$x$ is a defining property or parameter}
            }{
                \sem{
                    x
                }_{S, N, R} =
                name(N $+\!\!+$ [x])
            }
            \quad(\textsc{DefProp-Ref/Param-Ref})
        \end{mathpar}

        \vspace*{-1.5\baselineskip}

        \begin{mathpar}
            \inferrule{
                \text{$x$ is a location} \\
                x \rightarrow x' \notin R
            }{
                \sem{
                    x
                }_{S, N, R} =
                name(N $+\!\!+$ [x])
            }
            \quad(\textsc{Location-Ref})
        \end{mathpar}

        \vspace*{-1.5\baselineskip}

        \begin{mathpar}
            \inferrule{
                \text{$x$ is a location} \\
                x \rightarrow x' \in R
            }{
                \sem{
                    x
                }_{S, N, R} =
                name(N $+\!\!+$ [x'])
            }
            \quad(\textsc{ActionLocation-Ref})
        \end{mathpar}

        \vspace*{-1.5\baselineskip}

        \begin{mathpar}
            \inferrule{
                \text{$x$ is a derived property with body $e$}
            }{
                \sem{
                    x
                }_{S, N, R} =
                \sem{
                    e
                }_{S, N, R}
            }
            \quad(\textsc{DerProp-Ref})
        \end{mathpar}

        \vspace*{-1.5\baselineskip}

        \begin{mathpar}
            \sem{
                e.x
            }_{S, N, R} = \sem{
                e
            }_{S, N, R} + \_x
            \quad(\textsc{Proj})
        \end{mathpar}
    \end{mdframed}

    \vspace*{-1.25\baselineskip}
    \caption{
        Denotational semantics of CSX, expressed in MiniZinc.
        We have omitted the rules for literals and arithmetic for brevity; they map one-to-one.
        $+\!\!+$ is namespace concatenation.
        $+$ is identifier concatenation.
    }

    \label{fig:denotational-semantics}
\end{figure*}

%% file: sections/05-implementation.tex
\section{Implementation}
\label{sec:implementation}

In this section we describe how we obtain a usable integrated development environment (IDE) for CSX by integrating an implementation of the language with configuration space exploration and interactive validation.
The IDE contains components for parsing, syntax highlighting, code completion, name binding and type checking, and interactive reporting of static semantics violations.
The CSX validation constructs are interpreted interactively and invalid assertions are marked on the specification.

We have implemented the CSX language using Spoofax~\cite{KatsV10}, a language workbench~\cite{ErdwegSVTBCGH0L15} that provides infrastructure for designing, implementing, and deploying DSLs by means of declarative specification of language aspects using meta-DSLs.
We define the syntax of CSX in SDF3~\cite{AmorimV20}, a meta-language for multi-purpose syntax definition.
From the CSX syntax definition, SDF3 automatically derives a parser, pretty printer, syntax highlighting, and syntactic code completion.
The parser yields abstract syntax trees (ASTs) on which we first apply desugaring.
Desugaring e.g. involves propagating the properties of a scenario to the tests within that scenario.
The desugared ASTs are input to the static analysis and further transformations.
We specify desugaring and other transformations using the Stratego~\cite{BravenboerKVV08} meta-language.
Based on the language specification, Spoofax automatically generates an IDE for the language.

We define the CSX static semantics in NaBL2~\cite{NeronTVW15, AntwerpenNTVW16}.
NaBL2 is a meta-language for specifying static semantics for languages from which name binding and type checking is automatically derived.
Static semantic violations are reported interactively in the IDE.
For CSX, this could be invalid composition of components in a device or incorrect type checking of constraint expressions.
Interactive reporting of errors assists users of the language during specification writing.

In addition to the automated derivation of name binding and type checking, we implement analysis for other well-formedness conditions.
If well-formedness checking succeeds, the result is a desugared AST that is annotated with name binding and typing information.
The name binding information is used to check non-cyclic references of defining properties and derived properties, i.e., by following references of properties and checking whether those do not contain cycles.

To realize configuration space exploration, we implement a translation of CSX specifications to SMT constraints for which we can use existing solving techniques.
In particular, we translate CSX to the MiniZinc constraint modelling language~\cite{NethercoteSBBDT07, StuckeyFSTF14}.
MiniZinc is solver-independent, which enables us to use multiple solvers as a backend for CSX.
In particular, we use solvers with the theories of linear arithmetic and optimization modulo theories.

We implement the translation from CSX to MiniZinc as an AST-to-AST transformation using Stratego.
In addition to the syntax definition of CSX, we have also defined the syntax of MiniZinc in Spoofax with SDF3\footnote{\url{https://github.com/metaborgcube/metaborg-minizinc}}.
The syntax definitions of both languages generate an AST schema on which we define the Stratego transformation.
After transforming a parsed CSX AST to a MiniZinc AST, the MiniZinc pretty printer generates concrete MiniZinc syntax from the AST.

The translation uses information from name binding and type analysis.
This is necessary for references and projection expressions.
By using name binding and typing information, the distinction between references to properties, parameters, locations, and action locations can be made to generate the correct reference on the MiniZinc level.

We integrate solving of constraint models by calling MiniZinc from Stratego through integration with Java.
Stratego provides an API for integrating transformations with custom Java code.
We implement such a custom transformation and use a Java program to call the MiniZinc command-line interface.
The Java program is called with as input the generated MiniZinc model.
The Java program parses the textual solving result that is returned by MiniZinc and returns it as a list of variable binding.
In the Stratego code, for the interpretation of configurations, we evaluate expressions and lookup values for references by following the same naming schema as in the translation semantics.
After replacing the referenced properties and parameters by their values on the constraint level, the evaluation of expressions remains regular expression evaluation.
As a result, we have a configuration space exploration pipeline from interpreting specifications using constraint solving with the solution mapped back to the specification level as a configuration.

The configuration space exploration pipeline serves two purposes in the IDE: test evaluation and inhabitance checking.
For test evaluation, the configuration space of the device that is selected in the scenario is translated to MiniZinc and passed as an input to the pipeline.
Additional constraints are added to reduce the configuration space, e.g. to configure the input or output location values, or parameters as specified in the scenario.
If the scenario contains an objective, the objective is also mapped to MiniZinc and provided as input to the pipeline.
The configuration that is returned by the pipeline is used to evaluate test expectations.
This evaluation is done by a basic interpreter that evaluates expressions which should result into true.
The expressions can contain references to parameters and location properties, and based on the name binding information the references are mapped to the corresponding value from the configuration.
For failed test expectations we report an error which is marked with red underlining on the original specification using origin tracking~\cite{DeursenKT93}.

The evaluation of tests and reporting of results is triggered in the IDE on file changes, resulting into an interactive experience.
Additionally, the experience is improved by providing information while hovering over references to locations, properties, and parameters in test expectations.
The same interpretation approach as for test expectations is used to evaluate the expression being hovered over and the value is presented in a popup, giving the user insight in the configuration that is found.

Similar to the treatment of scenarios, inhabitance checks are triggered on file changes.
The pipeline is triggered for each type, action, and device using the translations semantics.
For inhabitance checking of a type, we translate a random instance of that type to SMT.
For an action, we instantiate it with instances for all its parameters.
Instead of finding a configuration for it, for inhabitance checking we only check satisfiability on the constraint level.
If the pipeline concludes insatisfiabilty, we report an error on the corresponding construct to indcate that the construct is not inhabited.

\input{figures/architecture-global}

To prevent unnecessary checking of inhabitance and evaluation of tests, we use simple caching of analysis results with ASTs of the subjects as the caching key.
If a type definition AST has not changed, it does not have to be checked again for inhabitance.
If a scenario has not changed, it does not have to be evaluated again.

While we have described the realization of a programming environment for CSX specifications, the eventual goal of CSX is to deploy control software to finishers.
Figure~\ref{fig:architecture} gives an overview of how configuration space exploration with CSX would with fit in a realistic setting.
The configuration space exploration component would be integrated with a software component, implemented using a general-purpose language, that provides a UI and that instructs low-level embedded software components.

%% file: figures/architecture-global.tex
\begin{figure}[t]
	\centering
	\includegraphics[width=0.7\columnwidth]{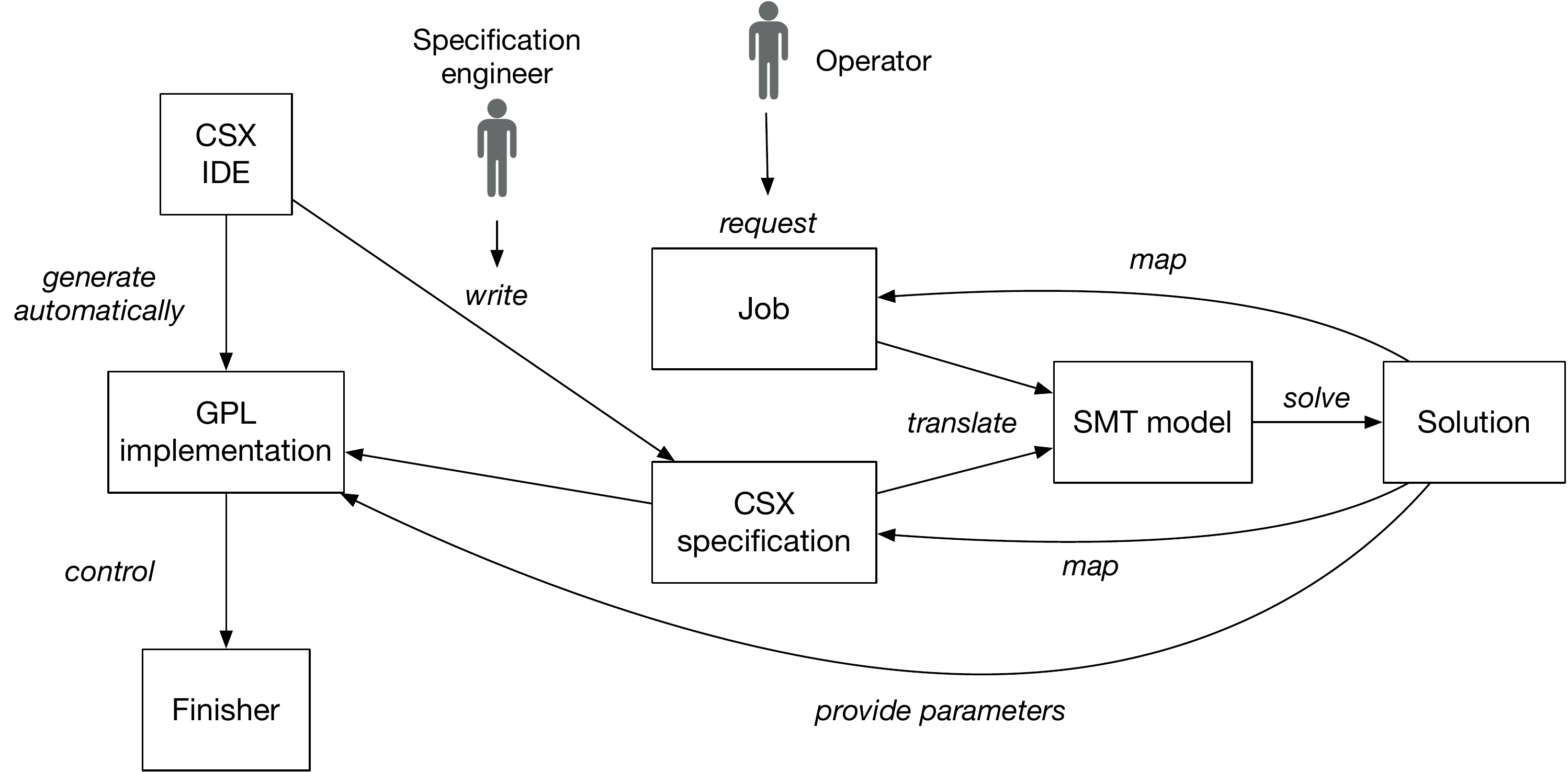}
	\caption{
		An architecture for applying CSX in control software.
		GPL stands for general purpose programming language, such as C\# or Java.
	}
	\label{fig:architecture}
\end{figure}

%% file: sections/06-evaluation.tex
\section{Evaluation}
\label{sec:evaluation}

We evaluate CSX by modelling two realistic cases, a perfect binder and a booklet maker, and by benchmarking the configuration space exploration for a scenario with and without optimization.
The perfect binder case corresponds to the example of Section~\ref{sec:csx}.
In the scenario without optimization, CSX derives the required input cover given an input bookblock and a desired output.
In the scenario with optimization, CSX finds a configuraton for the smallest size book the finisher can produce.
The bookletmaker case concerns a finisher that performs rotating, stitching, folding, and trimming in order to produce a booklet from a stack of sheets.
In the scenario without optimization, CSX finds the action parameters given an input and output.
In the scenario with optimization, CSX finds a configuration that minimizes paper waste given only the desired output.
Both specifications are based on realistic cases present at Canon Production Printing~B.V.

By writing scenarios in the language, we can interactively validate the specification within the IDE.
Initially loading a specification can take a few seconds: a specification typically consists of multiple type definitions, action definitions, a device definition, and several scenarios.
For the type, action, and device definitions, inhabitance checking is triggered, which for each check leads to an invocation of the SMT solver.
Additionally, for each scenario the solver is invoked.
The caching of invocations of the solver decreases response times after a change, making the IDE usable in an interactive way.
For example, inhabitance for a type will not be re-checked if only a test scenario changes.

\input{figures/benchmarks}

We set up a benchmark which makes use of Spoofax core, i.e. the core of Spoofax which enables integration of language components with Java, such that we can only execute the relevant part of the pipeline in the benchmark.
For benchmarking, we use the JMH framework\footnote{\url{https://openjdk.java.net/projects/code-tools/jmh/}}.
We executed the benchmarks on a server with two 32-core processors with a base frequency of 2.3GHz and 256GB RAM, running Ubuntu 20.04, using OpenJDK version 1.8.0\_275-b01.
From experimentation it appeared that the ORTools solver\footnote{\url{https://developers.google.com/optimization}} had best performance, and therefore we use this solver in the benchmarks. 
We use MiniZinc version 2.5.5 and ORTools version 9.0.
We measure 10 iterations and average the result.
In the benchmarks, we separately measure the translation time and solving time.
We leave out parsing, name binding and type checking time, as they are minimal compared to translation and solving time.

Figure~\ref{fig:benchmarks} shows the benchmarking results.
For each scenario, solving time is in the order of $100$'s milliseconds.
We consider sub-second performance as practical and therefore conclude that CSX's performance for the two cases we consider has practical performance for finding (optimal) configurations.

For specifying these devices in CSX, we have chosen a model of objects (sheets, stacks) with a certain level of detail.
The bookletmaker and perfect binding cases translated in the SMT level into 32 and 29 variables and 56 and 58 constraints, respectively.
Although we achieve useful configuration space exploration for these scenarios, it could be that in practice more detail has to be added to the model, which could also influence solving performance.
By deploying CSX at Canon Production Printing~B.V., we aim to further evaluate whether CSX is adequate in modeling and integrating the full product line of finishers available and evaluate its usability for domain experts.

%% file: figures/benchmarks.tex
\begin{figure}
	\centering
	\includegraphics[width=0.55\columnwidth]{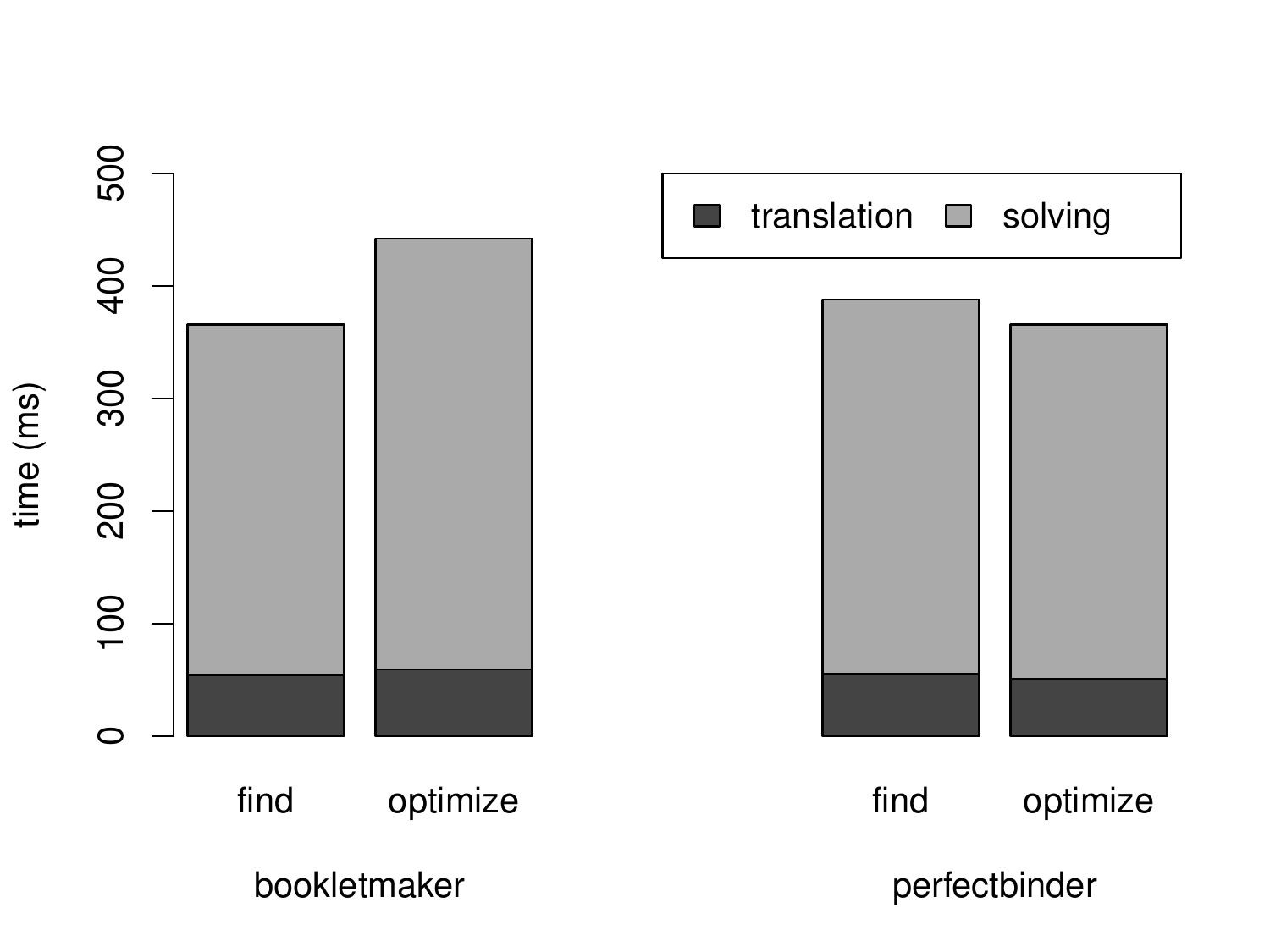}
	\caption{
		The benchmarking results on a perfect binder and a booklet maker for a scenario of finding a configuration and for finding an optimal configuration.
	}
	\label{fig:benchmarks}
\end{figure}

%% file: sections/07-related.tex
\section{Related Work}
\label{sec:related}

We discuss related work that uses constraint solving in the backend of high-level specification or domain-specific languages for realizing static analyses, validation, verification, consistency checking or synthesis.

Keshishzadeh et al. use SMT solving for validation of domain-specific properties to achieve fault detection early in the software development cycle.
In particular, they develop a DSL with industrial application in a case on collision prevention for medical imaging equipment~\cite{KeshishzadehMM13}.
The approach includes delta debugging, i.e., an approach to trace causes of property violations and report them back to the specification in a systematic way.
The work is related to CSX because it also uses SMT solving in the backend of a domain-specific language.

Voelter et al. use SMT solving with the Z3 solver for advanced error checking and verification in the KernelF language~\cite{Voelter18}, a reusable functional language for the development of DSLs.
Voelter et al. apply SMT solving succesfully in a DSL on a case study for the domain of payroll calculations~\cite{Voelter2020}, i.e. for statically checking completeness and overlap of domain-specific switch-like expressions.
Similarly to CSX, in this work SMT solving is used in the backed of a domain-specific language for realizing static analyses.
While the application of SMT was successful in the domain-specific case, the authors report difficulties in applying SMT solving generically in KernelF.
The authors plan to develop a successor to KernelF that is realized with SMT solving completely.

Constraint solving in feature models solves a different problem than CSX.
Feature models describe systems as compatible compositions of features or software components; finding/checking feature compositions occurs ``statically'' from which a software artifact can be derived.
CSX specifications express physical properties of finishers; finding configurations occurs ``dynamically'' (at run time) to find instances of the manufacturing process. This goes all the way down to the ``semantic'' level, e.g. by using sheet dimensions and the location of fold edges instead of only an abstract feature that enumerates the kinds of folds a device can do.
Feature modelling is useful in the finishing context e.g. to derive which devices are necessary for a production route for booklets. In CSX, we assume the production route is known.

Relational model finders are related to CSX in the sense that they map high-level specifications to constraints and map solutions back to the specification level.
Alloy~\cite{Jackson02-8} is a specification language that applies finite model finding to check formal specifications of software.
Alloy is backed by KodKod~\cite{TorlakJ07}, a relational model finder for problems expressed using first order logic, relational algebra, and transitive closures.
In contrast to CSX, KodKod does not offer support for reasoning over data nor for optimization objectives. 
In CSX, the nature of specifications is not relational: manufacturing paths are fixed and we consider snapshots of the product being manufactured at different steps in the process.

AlleAlle~\cite{StoelSV19} adds support for first-class data attributes and optimization to relational model finding.
Similar to KodKod, Stoel et al. consider AlleAlle as an intermediate language.
AlleAlle and CSX are related in the sense that both approaches take the data of problems into account and use SMT solving for model finding.
While AlleAlle is an intermediate language generally targeting relational problems, CSX is a more domain-specific language in which relations are not a first class concept.
Similar to CSX, for AlleAlle it is unclear yet how to map reasons for unsatisfiability that are found in the constraint level back to the specification level.

Rosette~\cite{TorlakB13} is a solver-aided programming language that supports verification, debugging, and synthesis.
Rosette extends the Racket language with support for symbolic values that stand for e.g. an arbitrary integer value.
Such values translate to a constraint variable in the runtime.
Rosette realizes verification and synthesis in the runtime by integrating its symbolic virtual machine with SMT solvers.
Whereas in Rosette selected variables are replaced by symbolic values, in CSX all variables in the specification translate to constraint variables.
Rosette is a general language tailored to program verification and synthesis whereas CSX is focused on a particular domain, i.e. manufacturing systems, altough we have only experimented with CSX in the digital printing domain.

Muli~\cite{DagefordeK19-0} is a constraint-logic object oriented language that integrates constraint solving with object oriented programming in the Java programming language.
Muli extends Java's syntax with the \texttt{free} keyword for indicating symbolic values that translate to constraint variables in the runtime.
Fragments of programs that are considered as search regions are executed non-deterministically, searching for concrete values for the constraint variables.
The Muli runtime is based on a symbolic Java virtual machine that integrates constraint solvers.
Muli only supports primitive types as constraint variables.
Support for arrays and objects as constraint variables is listed as future work.
CSX does support search on non-primitive types such as user-defined record types.
Similar to how support for arrays is desired for Muli, support for lists is desired for CSX, but that is future work.
Muli differs from CSX in the sense that Muli preserves the Java syntax and, by doing so, serves as a general purpose programming language, whereas CSX introduces a new domain-specific language.
In contrast to Muli, CSX supports optimization.

%% file: sections/08-conclusions.tex
\section{Conclusions}
\label{sec:conclusions}

We have presented CSX, a language and method for high-level declarative specification of finishers and their configuration spaces.
We have developed a translation of CSX to SMT constraints which enables us to use constraint solving to find (optimal) configurations for finishers.
We have presented an implementation of the CSX programming environment, including support for well-formedness checking, inhabitance checking, and interactive configuration space exploration.
Our benchmarks show that, on two realistic cases, CSX has practical sub-second performance in finding configurations for scenarios with and without optimization.

\emph{Future work.}
Our focus has been on finding a domain abstraction for configuration space exploration applied in the digital printing domain for finishers.
While we have designed the language in collaboration with control software engineers, we plan to further evaluate CSX by deploying it at Canon Production Printing~B.V.
By doing so, we can further evaluate the adequacy of CSX in covering the full product line of finishers.
Additionally, we plan to evaluate the language in terms of usability for control software engineers and in terms of validatability by mechanical engineers.

To improve the usability of the environments for configuration space exploration for operators, it would be useful to characterize the reduced configuration spaces for given jobs.
In particular, when multi-objective optimization is relevant for ojectives such as maximing throughput and minimizing waste, it would be useful if CSX could indicate the tradeoff between these objectives.

%% file: appendices.tex
\ifisdraft\clearpage\fi
\appendix

\input{sections/09-declarative-semantics}
\input{sections/10-inhabitance}

%% file: sections/09-declarative-semantics.tex
\section{Declarative Semantics}
\label{sec:declarative-semantics}

\input{figures/declarative-semantics}

Specifications in CSX describe configuration spaces of devices.
For a device specified in CSX, a configuration assigns values to the locations and parameters of the device.
A valid configuration is a configuration that satisfies all constraints of the device.
We describe the satisfiability relation of CSX by defining the declarative semantics of CSX in Figure~\ref{fig:declarative-semantics}.
The rules follow the same pattern as the rules of the denotational semantics in Figure~\ref{fig:denotational-semantics}.
The configuration space of a device corresponds to the set of all valid configurations that satisfy the declarative semantics.

%% file: figures/declarative-semantics.tex
\begin{figure*}

    \begin{mdframed}
        $M \models_{S,R} S'$ \hfill $M$ models specification part $S'$ of $S$ with location renaming $R$ \\
        $M \models_{S,R} S' \Rightarrow v$ \hfill $M$ models specification part $S'$ of $S$ and evaluates to value $v$ \\
        \hphantom{x} \hfill with location renaming $R$ \\
        $R = \{ \dots, L_i \rightarrow L'_i, \dots \}$ \hfill Renaming of location names $L_i$ to $L'_i$ \\
        $D(T)$ \hfill Domain of type $T$. $D(\texttt{bool}) = \{\top, \bot\}$, $D(\texttt{int}) = \mathbb{Z}$. \\
        Locations $L$, components $C$, constraints $E$, defining properties $P$, types $T$, action parameters $PM$.
    \end{mdframed}

    \vspace*{-1.5\baselineskip}

    \begin{mdframed}[frametitle={Devices}]
        \vspace*{-0.5\baselineskip}
        \begin{mathpar}
        
            \inferrule{
                M \models_{S, \emptyset} L_1 \dots L_n \\
                M \models_{S, \emptyset} C_1 \dots C_m \\
                M \models_{S, \emptyset} E_1 \dots E_q
            }{
                M \models_{S, \emptyset} \texttt{device } d \texttt{ \{ }
                    L_1 \dots L_n,
                    C_1 \dots C_m,
                    E_1 \dots E_q,
                    \dots
                \texttt{ \}}
            }
            \quad (\textsc {Device})
        
        \end{mathpar}
    \end{mdframed}

    \vspace*{-1.5\baselineskip}

    \begin{mdframed}[frametitle={Locations}]
        \vspace*{-0.5\baselineskip}
        \begin{mathpar}

            \inferrule{
                \texttt{type } T \texttt{ \{ }
                    P_1 \texttt{:} T_1 \dots P_n \texttt{:} T_n,
                    E_1 \dots E_m,
                    \dots
                \texttt{ \}} \in S \\
                M.L = v \\
                v \models_{S, \emptyset} P_1 \texttt{:} T_1 \dots P_n \texttt{:} T_n \\
                v \models_{S, \emptyset} E_1 \dots E_m
            }{
                M \models_{S, \emptyset} \texttt{location } L : T
            }
            \quad (\textsc {Location})
    
            \inferrule{
                T \in \{ \texttt{int}, \texttt{bool} \} \\
                M.P \in D(T)
            }{
                M \models_{S,\emptyset} P \texttt{ : } T
            }
            \quad (\textsc {DefProp-PrimType})
    
            \inferrule{
                \texttt{type } T \texttt{ \{ }
                    P_1 \texttt{:} T_1 \dots P_n \texttt{:} T_n,
                    E_1 \dots E_m,
                    \dots
                \texttt{ \}} \in S \\
                M.P = v \\
                v \models_{S, \emptyset} P_1 \texttt{:} T_1 \dots P_n \texttt{:} T_n \\
                v \models_{S, \emptyset} E_1 \dots E_m
            }{
                M \models_{S,\emptyset} P \texttt{ : } T
            }
            \quad (\textsc {DefProp-DefType})
        
        \end{mathpar}
    \end{mdframed}

    \vspace*{-1.5\baselineskip}

    \begin{mdframed}[frametitle={Components}]
        \vspace*{-1.5\baselineskip}
        \begin{mathpar}

            \inferrule{
                \texttt{action } A \texttt{(}
                    L_1 \texttt{:} T_1^L \dots L_n \texttt{:} T_n^L
                \texttt{)} \\ \texttt{\{}
                    \texttt{parameter } PM_1 : T_1^P \dots \texttt{parameter } PM_m : T_m^P,
                    E_1^A \dots E_q^A,
                    \dots
                \texttt{\}} \in S \\
                R = \{ L_1 \rightarrow L'_1, \dots, L_n \rightarrow L'_r \} \\
                M.C = v \\\\
                v \models_{S, \emptyset} \texttt{parameter } PM_1 : T_1^P \dots \texttt{parameter } PM_m : T_m^P \\
                v \models_{S, R} E_1^A \dots E_q^A \\
                v \models_{S, \emptyset} E_1^C \dots E_s^C
            }{
                M \models_{S, \emptyset} \texttt{component } C \texttt{ = } A \texttt{ ( }
                    L'_1 \dots L'_r
                \texttt{ ) \{ }
                    E_1^C \dots E_s^C
                \texttt{ \}}
            }
            \quad (\textsc {Comp})

            \inferrule{
                T \in \{ \texttt{int}, \texttt{bool} \} \\
                M.PM \in D(T)
            }{
                M \models_{S, \emptyset} \texttt{parameter } PM \texttt{ : } T
            }
            \quad (\textsc {Param})
        
        \end{mathpar}
    \end{mdframed}

    \caption{
        Declarative semantics of CSX (continued on next page).
        We have omitted the rules for literals and arithmetic for brevity; they map one-to-one.
        We define models $M$ recursively as $(M, x = v)$ in which value $v$ binds to $x$ and with the empty model $\varnothing$ as base case.
        We define model projection as $(M, x = v).x = v$ and $(M, x = v).y = M.y$ if $x \neq y$.
        $e \Rightarrow v$ indicates that syntactic expression $e$ evaluates to value $v$.
        Values are booleans ($\top$ and $\bot$), integers, and models.
    }

    \label{fig:declarative-semantics}
\end{figure*}

\begin{figure*}
    \ContinuedFloat

    \vspace*{-1.5\baselineskip}

    \begin{mdframed}[frametitle={Constraints \& References}]
        \vspace*{-0.5\baselineskip}
        \begin{mathpar}

            \inferrule{
                M \models_{S, R} e \Rightarrow v \\
                v = \top
            }{
                M \models_{S, R} \texttt{[ } e \texttt{ ]}
            }
            \quad (\textsc {Constraint})
    
            \inferrule{
                \text{$x$ is a defining property or parameter} \\
                M.x = v
            }{
                M \models_{S,R} x \Rightarrow v
            }
            \quad (\textsc {DefProp-Ref/Param-Ref})
    
            \inferrule{
                \text{$x$ is a location} \\
                x \rightarrow x' \notin R \\
                M.x = v
            }{
                M \models_{S,R} x \Rightarrow v
            }
            \quad (\textsc {Location-Ref})
    
            \inferrule{
                \text{$x$ is a location} \\
                x \rightarrow x' \in R \\
                M.x' = v
            }{
                M \models_{S,R} x \Rightarrow v
            }
            \quad (\textsc {ActionLocation-Ref})

            \inferrule{
                \text{$x$ is a derived property with body $e$} \\
                M \models_{S,R} e \Rightarrow v
            }{
                M \models_{S,R} x \Rightarrow v
            }
            \quad (\textsc {DerProp-Ref})
    
            \inferrule{
                M \models_{S,R} e \Rightarrow v1 \\ v1.x = v2
            }{
                M \models_{S,R} e.x \Rightarrow v2
            }
            \quad (\textsc {Proj})
        
        \end{mathpar}
    \end{mdframed}
 
    \caption{
        Declarative semantics of CSX (continued).
    }

    \label{fig:declarative-semantics}
\end{figure*}

%% file: sections/10-inhabitance.tex
\section{Inhabitance}
\label{sec:inhabitance}

\input{figures/denotational-semantics-inhabitance}

The CSX syntax allows to define types, actions, and devices without inhabitants.
For example, the following type is not inhabited:

\begin{lstlisting}[language=csx]
type T { i: int [i != i] }
\end{lstlisting}

Since there are no valid configurations for such definitions, we want to detect and report uninhabited definitions.
Specifications with definitions that are not inhabited, i.e., there are no models for their instantations, are not useful in practice.
Therefore, we restrict CSX such that types, actions, and devices must be inhabited.
Below, we define inhabitance in terms of the satisfiablity relation of Figure~\ref{fig:declarative-semantics}.

A type $T$ is inhabited if there exists a model $M$ that is a value for a arbitrary location $L$ of type $T$ and that satisfies the specification of the type:

\begin{mdframed}
\[
    \inferrule{
        \text{arbitrary name $L$}
    }{
        \exists M \models_{\emptyset,\emptyset} \texttt{location } L \texttt{ : } T
    }
\]
\end{mdframed}

An action $A$ is inhabited if there exists a model $M$ that satisfies an instance of the action, i.e., a valuation its for parameters and that satisfies the specification of the action:

\begin{mdframed}
\[
    \inferrule{
        M \models_{\emptyset, \emptyset} \texttt{location } L_1 : T_1^L \dots \texttt{location } T_n : T_n^L \\
        R = \{ L_1 \rightarrow L_1, \dots, L_n \rightarrow L_n \} \\
        M \models_{\emptyset, \emptyset} \texttt{parameter } PM_1 : T_1^P \dots \texttt{parameter } PM_m : T_m^P \\
        M \models_{\emptyset, R} E_1^A \dots E_q^A
    }{
        \exists M \models_{\emptyset,\emptyset} \texttt{action } A \texttt{(}
            L_1 \texttt{:} T_1^L \dots L_n \texttt{:} T_n^L
        \texttt{)} \\ \texttt{\{}
            \texttt{parameter } PM_1 : T_1^P \dots \texttt{parameter } PM_m : T_m^P,
            E_1^A \dots E_q^A,
            \dots
        \texttt{\}} \\
    }
\]
\end{mdframed}

A device $d$ is inhabited if there exists a model $M$ that satisfies the device: $\exists M \models_{S,\emptyset} \texttt{device } d \texttt{ \{ } ... \texttt{ \}}$.

Inhabitance corresponds to satisfiability in the SMT domain.
Unsatisfiability of an SMT model for a type, action, or device indicates the definition is not inhabited.
For a device, it means the configuration space is empty.
The denotational semantics in Figure~\ref{fig:denotational-semantics} defines a translation from the perspective of a device.
We build on this translation to define rules for checking inhabitance of types and actions in Figure~\ref{fig:denotational-semantics-inhabitance}.

For inhabitance checking of types and actions, we can reuse the rules but have to provide an artifical context for the translation.
For example, for inhabitance checking of type $T$, we check satisfiability of the SMT model for an instance of the type in a arbitrary location (\textsc{Location-Inhab}).
The type is inhabited if the SMT model for an instance of the type in an arbitrary location is satisfiable.
For inhabitance checking of actions, we take a similar approach (\textsc{Action-Inhab}).
Instead of taking a single arbitrary location, we instantiate locations for all location paramaters, and use them to instantiate the action $A$ for an arbitrary component $C$.

%% file: figures/denotational-semantics-inhabitance.tex
\begin{figure*}

    \begin{mdframed}
        \vspace*{-\baselineskip}
        \begin{mathpar}
        
            \inferrule{
                \text{arbitrary name $L$}
            }{
                \sem {
                    \texttt{location } L : T
                }_{S, [], \emptyset}
            }
            \quad(\textsc{Location-Inhab})

            \inferrule{
                \texttt{action } A \texttt{(}
                    L_1 \texttt{:} T_1^L \dots L_n \texttt{:} T_n^L
                \texttt{)} \texttt{ \{} \\
                    \texttt{parameter } PM_1 : T_1^P \dots \texttt{parameter } PM_m : T_m^P,  \\\\
                    E_1^A \dots E_q^A,
                    \dots
                \texttt{ \}} \in S \\
                R = \{ L_1 \rightarrow L_1, \dots, L_n \rightarrow L_n \} \\\\
                \text{arbitrary name $C$}
            }{
                \bigcup_{i=1}^{n}
                \sem { \texttt{location } L_i : T_i^L }_{S, [], \emptyset}
                \cup \\\\
                \bigcup_{i=1}^{m}
                \sem { \texttt{parameter } PM_m : T_m^P }_{S, [C], \emptyset}
                \cup \\\\
                \bigcup_{i=1}^{q}
                \sem {
                    E_i^A
                }_{S, [C], R}
            }
            \quad(\textsc{Action-Inhab})
        
        \end{mathpar}
    \end{mdframed}

    \caption{
        Denotational semantics for inhabitance checking, building on the rules of Figure~\ref{fig:denotational-semantics}.
    }

    \label{fig:denotational-semantics-inhabitance}

\end{figure*}